\documentclass[aps,prl,reprint,superscriptaddress,longbibliography]{revtex4-2}
\usepackage{graphicx}
\usepackage{color}
\usepackage[tight]{units}
\usepackage{epstopdf}
\usepackage[normalem]{ulem}
\DeclareGraphicsExtensions{.pdf,.jpg,.png,.eps}
\usepackage{amsfonts}
\usepackage{amsmath}
\usepackage{amssymb}
\usepackage{color}
\usepackage{color}
\usepackage{lineno}
\usepackage{epstopdf}
\usepackage{adjustbox}

\usepackage[T1]{fontenc}
\usepackage{hyperref}
\newcommand{\sample}{\ensuremath{\mathrm{Nb}_{0.45}\mathrm{Ta}_{0.25}\mathrm{Ti}_{0.15}\mathrm{Hf}_{0.15}}}
\begin{document}
\title{Interplay of Chemical Disorder, Vortex Pinning, and Electron Correlations in the Mechanically Robust Medium-Entropy Alloy Superconductor \sample}

\author{Pooja Chourasia}
\affiliation{Department of Physics and Material Science, Thapar Institute of Engineering and Technology, Patiala 147004, India}
\author{Samiul Sk}
\affiliation{Department of Physics, Bennett University,Greater Noida,201310,Uttar Pradesh,India}
\author{Nikita Sharma}
\affiliation{Department of Physics and Material Science, Thapar Institute of Engineering and Technology, Patiala 147004, India}
\author{B. Vignolle}
\affiliation{Univ. Bordeaux, CNRS, Bordeaux INP, ICMCB, UMR 5026, Pessac, F-33600, France}
\author{Abhishek Sarkar}
\affiliation{Department of Materials Science and Engineering, Indian Institute of Technology Delhi,New Delhi, 110016, India},
            
\author{Olivier Toulemonde}
\affiliation{Univ. Bordeaux, CNRS, Bordeaux INP, ICMCB, UMR 5026, Pessac, F-33600, France}
\author{Sanjoy Mahatha}
\affiliation{UGC-DAE Consortium for Scientific Research, Khandwa Road, Indore 452001, Madhya Pradesh, India}
\author{S. K. Panda}
\affiliation{Department of Physics, Bennett University,Greater Noida,201310,Uttar Pradesh,India}
\author{Sourav Marik}
\email[]{soumarik@thapar.edu}
\affiliation{Department of Physics and Material Science, Thapar Institute of Engineering and Technology, Patiala 147004, India}
\begin{abstract}
Medium- and high-entropy alloys have emerged as a promising class of superconducting materials because their intrinsic chemical disorder can simultaneously enhance mechanical robustness and modify superconducting properties. Here, we present a combined experimental and first-principles investigation of the structural, superconducting, electronic, and local lattice-distortion characteristics of the mechanically robust medium-entropy alloy \sample. The material shows type-II superconductivity with a transition temperature $T_C^{onset} \approx 7.8 ~\mathrm{K}$. A high critical current density of $J_c = 0.48 \times 10^{6}~\mathrm{A/cm^2}$ is observed, and this is attributed to strong vortex pinning at grain boundaries. Specific-heat analysis indicates a strong electron-phonon coupling ($\lambda$ = 1.07) with a large isotropic s-wave superconducting gap. Density functional theory calculations within the generalized gradient approximation (GGA), employing a special quasirandom structure (SQS) to explicitly model the intrinsic chemical disorder, predict a dynamically stable body-centered cubic phase with an equilibrium lattice parameter in excellent agreement with experiment. These calculations further reveal a metallic electronic structure with substantial Ti- and Ta-derived states at the Fermi level. Additionally, our results quantify a pronounced disorder-induced local lattice distortion ($\Delta d = 0.086$~\AA), providing a microscopic basis for the experimentally observed enhanced vortex-pinning landscape. Remarkably, the alloy also exhibits an unusually large Kadowaki-Woods ratio, placing it close to the heavy-fermion regime and indicating enhanced electronic correlations beyond conventional alloy superconductors. Our results demonstrate that entropy-engineered chemical disorder can simultaneously stabilize robust superconductivity, promote strong flux pinning, and induce correlated electronic behavior, establishing refractory medium-entropy alloys as a promising platform for next-generation superconducting materials.

\end{abstract}
\maketitle
\section{Introduction}
Superconducting materials capable of sustaining high critical current densities while maintaining excellent structural and mechanical integrity are essential for next-generation technologies, including fusion reactors, high-field magnets, particle accelerators, quantum technologies, and advanced power transmission systems \cite{1, 2, 3}. Although conventional superconductors have enabled many technological advances, achieving an optimal balance between superconducting performance, mechanical robustness, and environmental stability remains a persistent challenge. In this context, the emergence of complex entropy-stabilized alloys (high- and medium-entropy alloys) has transformed materials design by demonstrating that chemical complexity can yield combinations of properties that are difficult to achieve in conventional alloys \cite{4,5,6}. Unlike conventional alloys that rely on one or two principal elements, high entropy alloys (HEAs) and medium entropy alloys (MEAs) contain four or more elements mixed in equiatomic ratios \cite{5,7,8}. Thermodynamically, the stability of HEAs and MEAs anticipated by the high entropy of mixing according to the Gibbs free energy equation: \begin{equation*} \Delta G_{mix} = \Delta H_{mix} - T \Delta S_{mix} \end{equation*} where T is the temperature, $\Delta H_{mix} $ is enthalpy of mixing and $\Delta S_{mix}$ is configurational entropy \cite{9}. The configurational entropy increases logarithmically with an increase in constituent elements as  $\Delta S_\text{mix}^\text{conf} = -R \sum_{i=1}^N c_i \ln c_i$, where $R$ is the gas constant, $N$ the number of components, and $c_i$ the atomic fraction of the $i$-th element. At elevated temperatures, the resulting $T\Delta S_{mix}$ term becomes large enough to overrule the mixing enthalpy, thereby stabilizing simple solid-solution phases over intermetallic or complex phases according to Hume-Rothery criteria. Conventionally, alloys with $0.69R \leq \Delta S_{mix} < 1.61R$ are known as MEAs, whereas those with $\Delta S_{mix} \geq 1.61R$ are categorized as HEAs
\cite{10}. Most of the reported HEAs and MEAs crystallize in BCC, CsCl, HCP, and $\alpha$-Mn-type phases \cite{11,12,13}.

A defining characteristic of entropy-stabilized alloys is their exceptional mechanical performance, including high yield strength, high fracture toughness, superior microhardness, and remarkable fatigue resistance, which significantly outperform those of conventional alloys \cite{14}. Moreover, their high damage tolerance and structural stability under intense particle bombardment make them attractive candidate for application in the severe radiation and harsh environments associated with nuclear reactors \cite{15}.

These exceptional structural features, combined with the intrinsic chemical complexity of entropy-stabilized alloys, have sparked growing interest in exploring superconducting properties, motivated by the need for mechanically robust superconducting materials for operating under extreme conditions. The first superconductivity was discoverd in $\mathrm{Ta_{34}}\mathrm{Nb_{33}}\mathrm{Hf_{8}}\mathrm{Zr_{14}}\mathrm{Ti_{11}}$, which is a type-II superconductor with $T_C = 7.3 K$ and upper critical field = 8.2 T \cite{16}. This pioneering work established that superconductivity can persist despite extreme chemical disorder observed in these systems. Since then, several refractory HEAs and MEAs have been reported, focusing on exploring new materials with improved performance and the potential for unconventional superconductivity \cite{16,17,18,19}. For instance, an equimolar ScTiVNbCu HEA, which exhibits a high $J_c$ of 0.18 $\times10^6 A/cm^{2}$
 and a high upper critical field of $\mu_{0}H_{c2}$ of 12.9 T \cite{18}. In eutectic $\mathrm{Nb}\mathrm{Sc}\mathrm{Ti}\mathrm{Zr}$  $J_c$ also exceeds $10^{6} A/cm^2$ while in the BCC $\mathrm{Nb_{2/6}}\mathrm{Ta_{1/6}}\mathrm{Ti_{1/6}}\mathrm{Zr_{1/6}}\mathrm{Hf_{1/6}}$ $J_c$ is of the order of $10^{5} A/cm^2$ at 2 K \cite{20,21,22}. Thin films of $\mathrm{Nb}\mathrm{Ti}\mathrm{Ta}\mathrm{Hf}\mathrm{Zr}$ exhibit a high $J_c$ exceeding $10^{6} A/cm^2$ at 4.2 K \cite{23}. Eqiatomic $\mathrm{Nb}_{0.25}\mathrm{Ta}_{0.25}\mathrm{Ti}_{0.25}\mathrm{Zr}_{0.25}$ shows multiband feature with high $J_c$ value of $\approx 0.13 \times10^{6} A/cm^2$ \cite{19}. Also, Ti-rich HEAs have been reported to exhibit a high $\mu_{0}H_{c2}(0) = 13.5 T $ \cite{24}. \\

Recently, exceptional fracture toughness even at cryogenic temperature has been observed in MEA \sample \, \cite{25, 75}.This material highlights fracture toughness exceeding 253~MPa$\cdot$m$^{1/2}$ over a wide temperature range of 77--1473~K, along with room-temperature yield strength of 596~MPa. The remarkable deformation behavior in \sample{} is attributed to the kink band formation driven by the coordinated slip of $\langle111\rangle$ edge dislocations across intersecting $\{110\}$ and $\{112\}$ slip systems, while edge dislocation pile up on $\{112\}$ planes, initiates the dislocation motion that causes kink band formation. The repeated nucleation and propagation of these kink bands accommodate heterogeneous deformation, redistribute local stresses away from crack tips, and mitigate strain localization, thereby establishing an intrinsic micromechanical hardening mechanism and enabling exceptional ductility and fracture resistance in this refractory BCC MEA \cite{25}.\\

Motivated by the need for mechanically robust superconducting materials, we investigate the superconducting properties and local distortion effects of mechanically robust \sample. In entropy-stabilized alloys, the observed severe lattice distortion, spatial fluctuations in the local atomic environment, and associated defects not only enhance mechanical robustness but can also serve as effective vortex-pinning centers, thereby suppressing vortex motion and improving the critical current density. In this study, we synthesized the MEA \sample{} and conducted a systematic experimental and theoretical investigation to understand the phase stability, normal- and superconducting-state properties, and lattice-distortion effects in this material.
\begin{figure*}[t]
\includegraphics[width=1\linewidth]{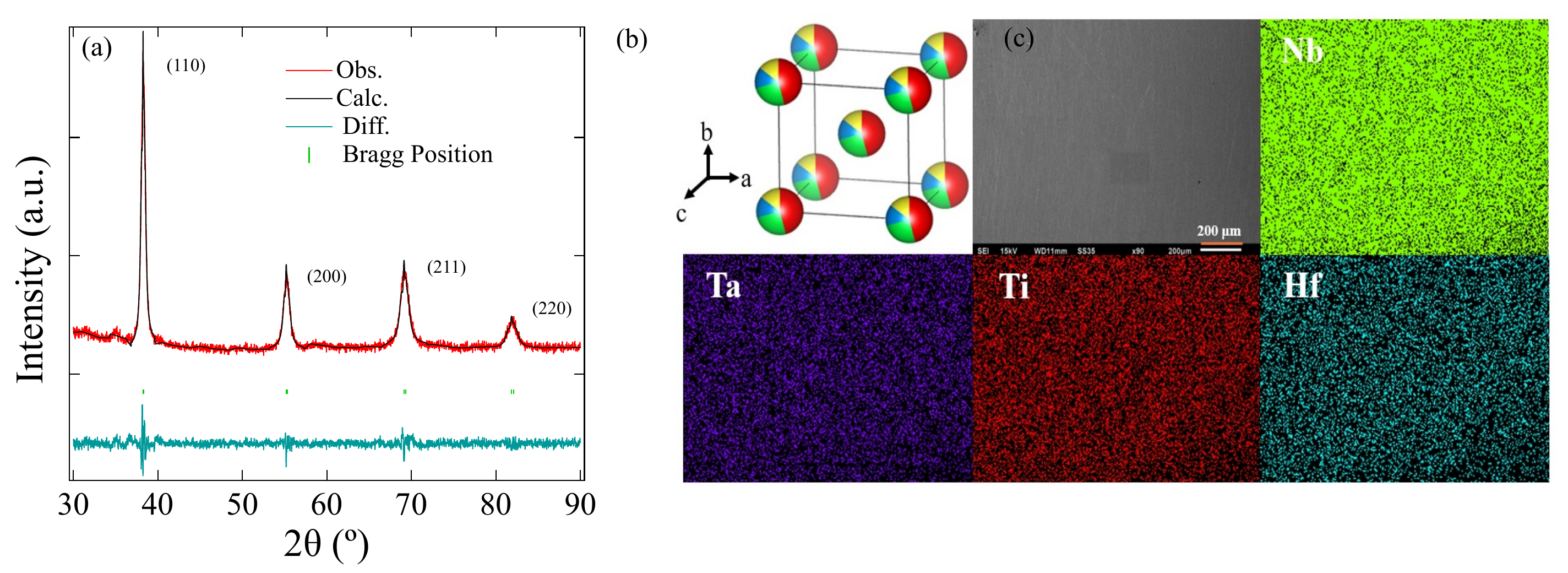}
 \caption{\label{Fig1:XRD}
    (a) The X-ray diffraction pattern of \sample{} at room temperature along with Rietveld refinement plot using Fullproof Suite. (b) Schematic representation of the BCC crystal structure, emphasizing the random occupation of lattice sites by the constituent elements in the \sample. 
    (c) FESEM micrograph along with EDS elemental maps for Nb, Ta, Ti, and Hf, illustrating the uniform distribution of the constituent elements throughout the alloy.}
\end{figure*}
\section{Methods}
\subsection{Experimental Methods }
\subsubsection{Sample Synthesis}
An entropy-stablized polycrystalline alloy of composition Nb$_{0.45}$Ta$_{0.25}$Ti$_{0.15}$Hf$_{0.15}$ was synthesized using the arc melting technique. The constituent elements were weighed in stoichiometric ratios and then melted in a high-purity argon atmosphere in the presence of a Ti getter. The alloy was melted to form an ingot, then flipped and remelted several times to ensure chemical homogeneity of the sample. The sample formed was hard and exhibits negligible weight loss after melting . 
\subsubsection{Structural and Microstructural Analysis } The structural and microstructural analysis of the alloy was done using a Rigaku X-Ray Diffractometer ( Rigaku SmartLab SE ) with Cu-K${\alpha 1}$ radiation ($\lambda = 1.5406$ \text{\AA}) and a ZEISS GEMINI field-emission scanning electron microscope (FESEM), respectively. Elemental distribution was determined using energy-dispersive X-ray spectroscopy (EDS). The Le Bail refinement of the resulting X-ray diffraction (XRD) patterns was performed using the FULLPROF Suite.
\subsubsection{Photoemission Spectroscopy}Photoemission spectroscopic measurements (PES), including X-ray Photo-electron spectroscopy (XPS), were performed on Thermoscientific Nexsa G2 series using a spot size of 400 $\mu\mathrm{m}$, whereas ultraviolet Photo-electron spectroscopy measurement (UPS) was done at AIPES beamline of the Indus-1 synchrotron radiation facility at Raja Ramanna Center for Advanced Technology (RRCAT), Indore,  using a photon energy of 40 eV at room temperature. Prior to photoemission measurements, the alloy surface was etched by using $\mathrm{Ar}^{+}$ ions at 3000 eV for 200 s to remove the oxide layer and surface contaminations. XPS provides information on core-level binding energies and oxidation states, while UPS provides valence-band features near the Fermi level. 
 
\subsubsection{Magnetic and Physical Properties} 
The magnetic properties were analyzed using a Superconducting Quantum Interference Device (SQUID) measurement system (MPMS-XL7).  The electrical transport and specific heat measurements were carried out using the Physical Property Measurement System (PPMS) by Quantum Design. The conventional four-probe method is used for electrical transport measurements.
\subsection{Computational Methods}
To provide a microscopic interpretation of the experimentally observed crystal structure, metallic conductivity, and stability of the MEA \sample{}, density functional theory (DFT)-based first-principles calculations were carried out~\cite{26,27}. Since the alloy exhibits intrinsic substitutional disorder arising from the random occupation of lattice sites by multiple constituent elements, an explicit treatment of chemical disorder is essential for obtaining a realistic description of its electronic structure. Hence, for a realistic description of this random occupation of the body-centered cubic (bcc) lattice while retaining periodic boundary conditions, we adopted the Special Quasirandom Structure (SQS)~\cite{28} approach. This approach is known to reproduce the most important correlation functions of an ideal random alloy and is therefore considered a well-established framework for investigating the electronic structure and lattice dynamics of chemically disordered high- and medium-entropy alloys. The SQS was generated using the \texttt{mcsqs} module of the Alloy Theoretic Automated Toolkit (ATAT), which employs a Monte Carlo simulated annealing algorithm to optimize the atomic configuration~\cite{29,30,31,32}. During the optimization, the atomic arrangement was chosen to reproduce as closely as possible the pair and higher-order correlation functions of an ideal random solid solution. A 20-atom supercell was found to provide an optimal balance between computational efficiency and an accurate representation of the experimentally observed substitutional disorder. The optimized SQS model was subsequently employed for all electronic-structure and lattice-dynamical calculations.\\
First-principles calculations were carried out based on the framework of density functional theory (DFT)~\cite{26,27} using the projector augmented-wave (PAW) method~\cite{33,34}, as implemented in the Vienna \textit{ab-initio} Simulation Package (VASP)~\cite{35,36,37}. The exchange-correlation effects were treated within the generalized gradient approximation (GGA)~\cite{38} using the Perdew-Burke-Ernzerhof (PBE) parametrization~\cite{39}. A plane-wave kinetic-energy cutoff of 550 eV was employed throughout the calculations to ensure well-converged total energies and electronic properties. Brillouin-zone integrations were carried out using an $8\times7\times7$ $\Gamma$-centered Monkhorst-Pack $k$-point mesh~\cite{40}. To obtain the equilibrium structure, both the lattice parameters and internal atomic coordinates were fully relaxed until the Hellmann-Feynman forces on every atom were smaller than $10^{-3}$ eV/\AA, while the electronic self-consistency was achieved with a total-energy convergence criterion of $10^{-6}$ eV. The optimized structure served as the starting point for all subsequent calculations. The calculated equilibrium lattice parameter was directly compared with the X-ray diffraction measurements, thereby providing an independent validation of the structural model adopted for the chemically disordered alloy.\\
 To assess the dynamical stability of the considered structure, phonon calculations were carried out within the framework of density functional perturbation theory (DFPT)~\cite{41,42,43,44} as implemented in VASP. The interatomic force constants and dynamical matrix were evaluated. The calculated phonon spectrum was further used to verify the absence of unstable vibrational modes, suggesting a dynamically stable phase of this structure. The calculations aim to establish a microscopic understanding of the experimentally observed structural and electronic properties by validating the stability of the chemically disordered bcc phase of \sample{} and elucidating the nature of the low-energy electronic states near the Fermi level. 
\section{Results and Discussion}
\subsection*{Structure and Microstruture}
Figure~\ref{Fig1:XRD}(a) represents the X-ray diffraction pattern of \sample{} collected at room temperature. The X-ray diffraction pattern of the sample confirms formation of a single phase alloy, with peak broadening due to the presence of disorder. The  Le Bail
refinement of recorded XRD \linebreak data performed using Fullproof Suite reveals a single phase body-centered cubic (BCC) crystal  structure corresponding to Im$\bar{3}$m space group. The lattice constant is estimated as $a=b=c=3.3257 (2) ~\text{\AA}$. Theoretically, the lattice parameter can be calculated by using Vegard's law of mixtures, which is given as $a_{\text{mix}} = \sum_i x_i a_i $,  where $ x_i$ and $a_i$ are the atomic fraction and lattice constant of the constituent elements present in the sample \cite{45}. The evaluated theoretical value of the lattice parameter is 3.34 \AA, close to the experimentally obtained value. Figure \ref{Fig1:XRD} (c) shows the FESEM image as well as the EDS mapping of the sample. The EDS mapping is performed at multiple positions across the sample to ensure chemical homogenity and the obtained average elemental composition is Nb$_{0.46}$Ta$_{0.27}$Ti$_{0.12}$Hf$_{0.16}$. The EDS data confirms the uniform distribution of all elements at the microscale in the sample. To assess the thermodynamic stability of the alloy, various phase formation parameters  are calculated and summarized in Table~I.
\subsection*{Photoelectron Spectroscopy}
\begin{figure*}[!t]
  \begin{flushleft}
      \includegraphics[width=\linewidth]{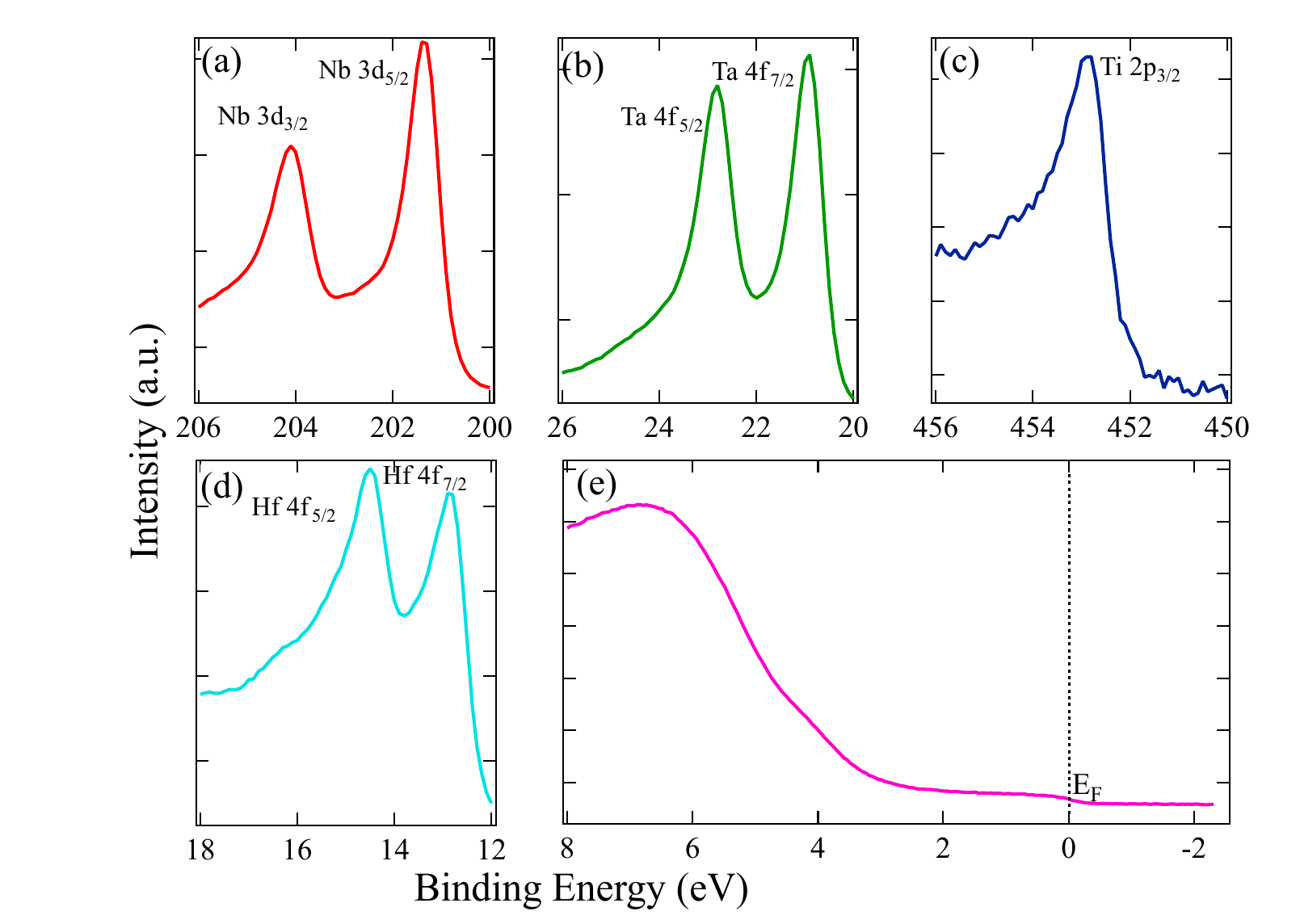}
    \end{flushleft}
    \caption{Photoemission spectra (PES) of core levels of (a) Nb 3d  (b) Ta 4f  (c) Ti 2p  (d) Hf 4f for HEA \sample and (black line) spectra for elemental metals at room temperature. (e) Valence band spectra obtained using $\mathrm{He\,II}$ at room temperature, showing the electronic states in the vicinity of the Fermi level $E_F$.} 
   \label{fig7:Photoemission spectra}  
\end{figure*}

To get a deeper insight into the electronic states and chemical environment of constituent elements in \sample, PES measurements were performed, using $\mathrm{Al\ K_{\alpha}}$ (1486.6 eV) for XPS and a photon energy of 40 eV at room temperature for UPS. The core level spectral regions corresponding to Nb $3d$, Ti $2p$, Hf $4f$, and Ta $4f$ are shown in Figure \ref{fig7:Photoemission spectra} (a-d). The core levels of Nb, Ta, and Hf exhibit spin-orbit split doublets showing asymmetric Doniach-Sunjic line shape, which shows the metallic nature of the alloy. This asymmetric Doniach-Sunjic line shape arises due to final state interaction of the core hole and the conduction electrons, causing a tail toward higher binding energies \cite{46,47}. However, Ti shows only a single peak at $2p_{3/2}$, but the Doniach-Sunjic type asymmetric
line shape confirms its metallic nature. The peak at  $2p_{1/2}$ is suppressed due to surface contamination or oxide layer formation.\\ 
The valence band UPS spectrum of the alloy, using a photon energy of 40 eV at room temperature, is shown in  Figure \ref{fig7:Photoemission spectra} (e). The spectral weight in the vicinity of the  Fermi level ($E_F$ ) was determined from the normalized intensity within the binding-energy range of 0--2~eV, providing a quantitative measure of the electronic states near $E_F$ \cite{48}. The spectrum exhibits finite intensity at $E_F$, with the corresponding spectral weight estimated to be approximately 0.71. To illustrate the overall valence-band electronic structure, the spectrum over the wider binding-energy range of 0–10~eV is also presented. The observed non-zero spectral weight at $E_F$ further confirms the metallic nature of the alloy.\\
\begin{figure*}[t]
\begin{flushleft}
\includegraphics[width=\linewidth]{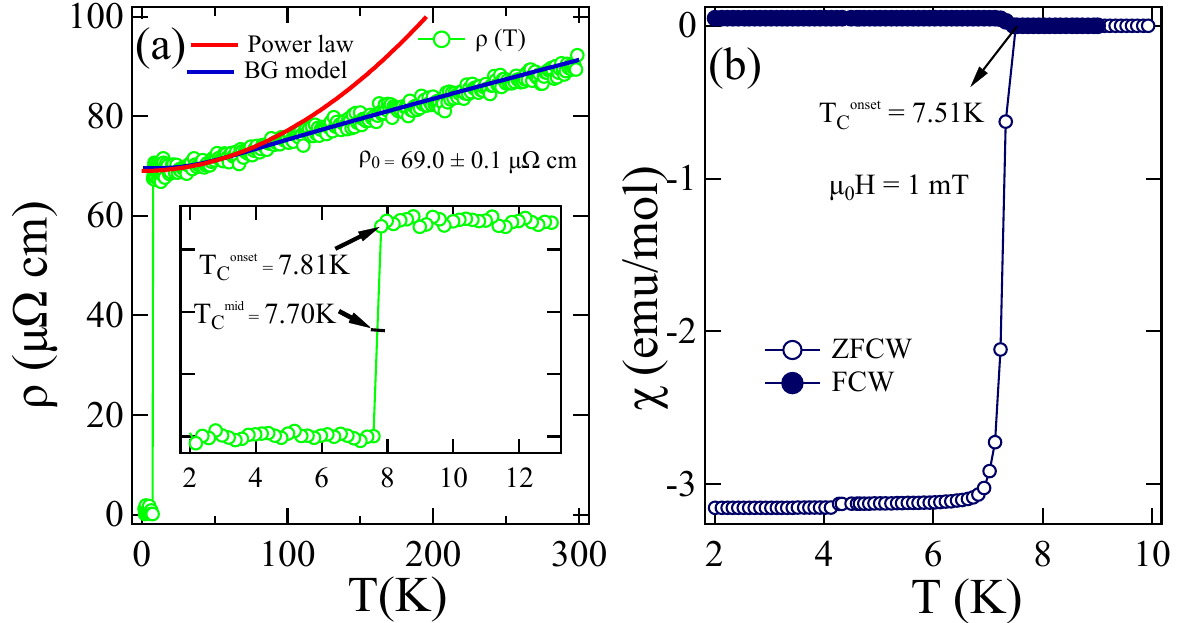}
 \end{flushleft}
 \caption{\label{Fig2:MT}
(a) DC electrical resistivity of \sample\ measured between 2 and 300 K in zero magnetic field, along with Bloch-Gruneisen  (blue) and Fermi liquid (red) model fits. The fitting was performed to extract the residual resistivity and elucidate the dominant scattering mechanisms. The inset indicates the superconducting onset temperature.  (b) Temperature dependent dc magnetic susceptibility measured under zero-field-cooled warming (ZFCW) and field-cooled warming (FCW) conditions in an applied magnetic field ($\mu_{0}H)$ of 1 mT.}
\end{figure*}
\begin{figure*}[htbp]
\hspace{-1 cm}
\includegraphics[width=1\linewidth,keepaspectratio]{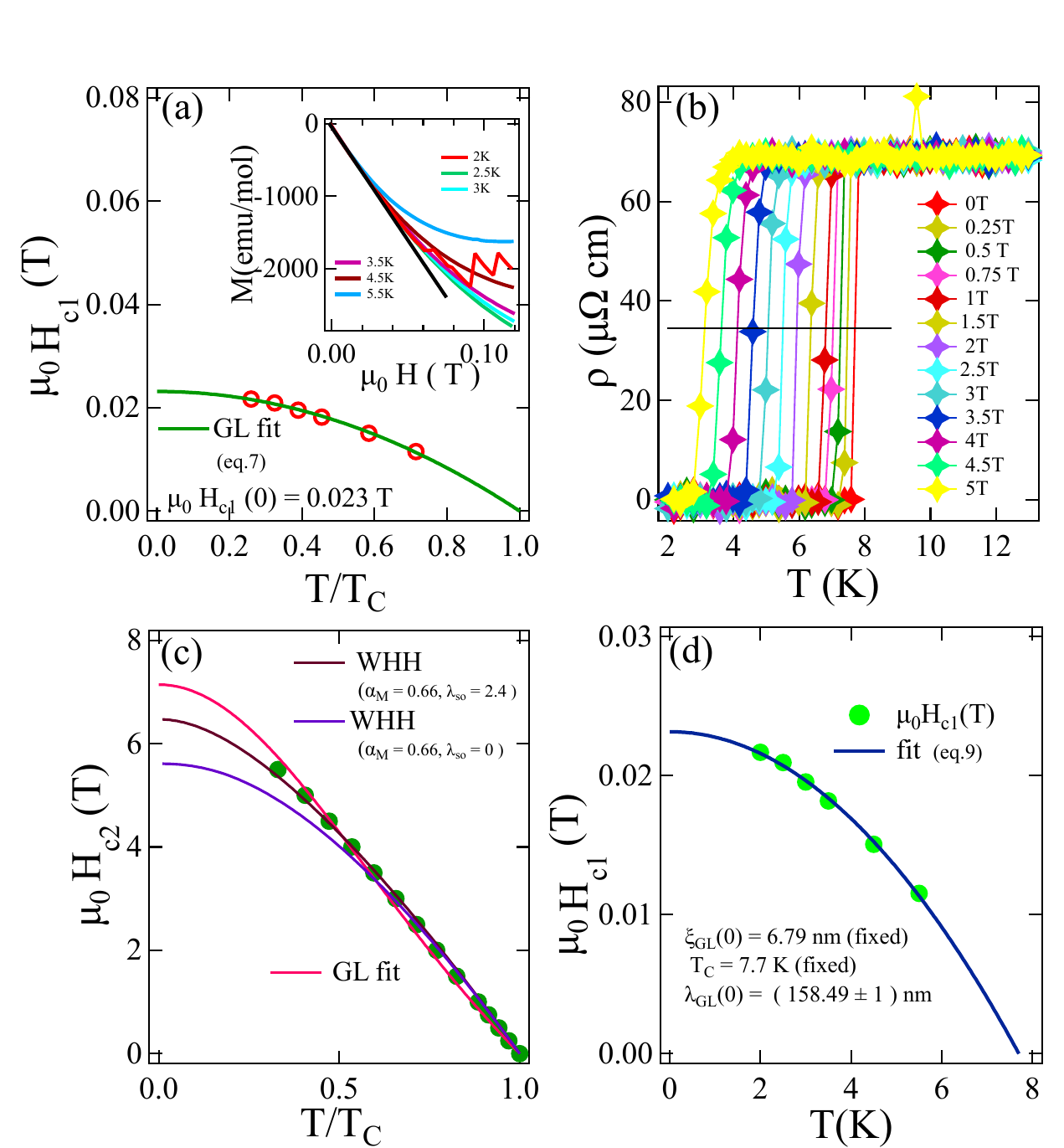}
\caption{\label{Fig3:Resistivity} (a) Temperature dependent lower critical field ($\mu_{0}H_{c1}(T)$) of the \sample{}, fitted using eq. \ref{eqn 3:hc1}. The inset shows magnetization curves (M-H) measured over the temperature range 2 K-5.5 K. (b) The field-dependent DC electrical resistivity as a function of temperature. (c) Temperature dependent upper critical field $\mu_{0}H_{c2}(T)$, solid line represents Ginzburg Landau (GL) fit. (d) The measured lower critical field $\mu_{0}H_{c1}(T)$ fitted as a function of temperature using eq. \ref{eq:Hc1fit}. The deduced penetration depth ( $\lambda_{GL} (0)$ ) from the fit (solid blue line) is 158.5 nm.}  
\end{figure*}
\subsection*{Normal and Superconducting-State Properties}
\begin{figure*}[!t]
\includegraphics[width=1\linewidth,keepaspectratio]{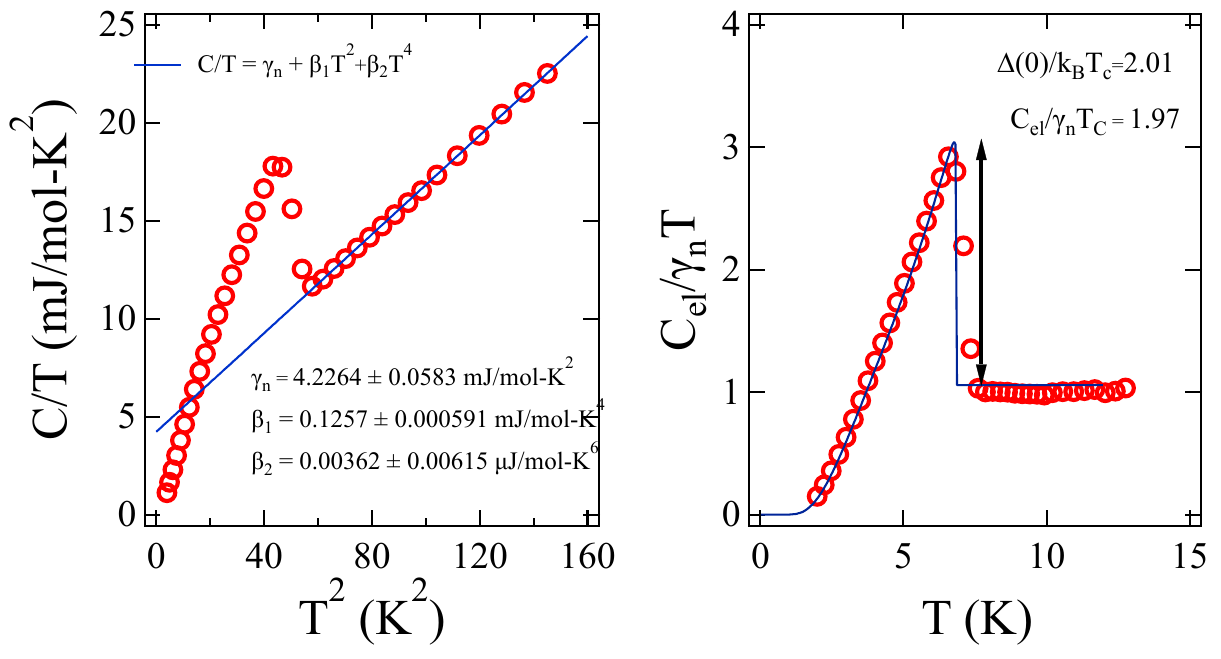}
    \caption{(a) C/T vs $T^2$ at zero field, fitted using eq.\ref{eqn 10:Debyemodel}  to extract electronic ($\gamma_n$) and phononic ($\beta$) and anharmonic contributions ($(\beta_2)$) contributions.  (b) The variation of electronic heat capacity with temperature, modeled using the isotropic s-wave model. The extracted superconducting gap ratio $\frac{\Delta(0)}{k_{B}T_{C}}$ equals to 2.01.}
    \label{fig5:Specific heat}  
\end{figure*}

The dc electrical resistivity ($\rho(T)$) as a fuction of temperature is shown in Figure~\ref{Fig2:MT} (a), is measured over 2 -- 300 K under zero applied magnetic field to study the superconducting as well as normal state properties of the alloy. The inset shows the transition to the superconducting state as resistivity begins to drop at $T_{c}^{onset}=7.81 K$. 
The residual resistivity ratio ($RRR$), given by $\rho_{300}/\rho_{10}$, is estimated to be 1.29. 
\begin{table}[ht!]
\caption{Compositional parameters for phase stability of \sample.}
\label{Parameters}
\centering
\small
\begin{tabular*}{1\columnwidth}{@{\extracolsep{\fill}}lcc}
\hline\hline
Parameters & Criteria (single phase) & \sample \\
\hline
$\Delta S_{mix}$ (J/K) & -- & $1.24$R \\
$\Delta H_{mix}$ (kJ/mol) & $\approx 0$ & $3.29$ \\
$\Omega$ & $1.1< \Omega <229.8$ & $78.03$ \\
$\Delta\chi$ & -- & $0.10$ \\
$\delta (\%)$ & $< 6.6$ & $4.31$ \\
VEC & $< 6.87$ (BCC) & $4.73$ \\
\hline\hline
\end{tabular*}
\end{table}
This low value of RRR indicates strong disorder-dominated scattering due to high atomic-scale disorder in \sample, comparable to other reported HEAs and MEAs \cite{18,49}. The characteristic quadratic temperature dependence of electrical resistivity is predicted by Fermi-liquid theory. The zero-field low temperature resistivity is well fitted using the Fermi liquid theory, which is given as :
\begin{equation}
    \rho = \rho_{0} + A T^n
    \label{eqn 1:Fermi liquid equation}
\end{equation}
where  $\rho _0$ is the residual resistivity resulting from temperature-independent scattering while A is the electron scattering coefficient extracted from the low-temperature resistivity region and n $=$ 2 for the Fermi-liquid behavior. The obtained values of $\rho_0$ and A are 69.016 $\mu\Omega\,\text{cm}$ and 0.00081279 $\mu\Omega\,\mathrm{cm}\,\mathrm{K}^{-2}$ respectively. The normal state resistivity of the sample was also fitted using the Bloch-Gruneisen (BG) model \cite{50,51}. The BG formula is given as:
\begin{equation}
\rho(T)=\rho_{0}+C\left(\frac{T}{\Theta_D}\right)^{n}
\int_{0}^{\Theta_D/T} 
\frac{x^{n}}{(e^{x}-1)(1-e^{-x})}\, dx.
\label{eqn 2:BG model}
\end{equation}
where C is the material-dependent constant, n identifies the primary dominant scattering mechanism in the material, and $\theta_D$ is the Debye temperature. The BG model best fitted using n=3, showing that electron-phonon scattering dominates the normal state electrical transport on this sample, and the obtained $\theta_D$ is 260 K. A similar trend is observed in various HEAs and MEAs   \cite{52,53}. Figure \ref{Fig3:Resistivity} (a) shows the temperature dependent dc electrical resistivity over magnetic field ranging from 0-- 6 T. The measurements were carried over the temperature range of 2 to 12 K. The upper critical field ($\mu_{0}Hc_{2}(0)$) is determined from temperature-dependent electrical resistivity. Figure \ref{Fig3:Resistivity} (c) shows temperature dependence of $\mu_{0}Hc_{2}(0)$ fitted using GL equation which is given as:
\begin{equation}
     \mu_{0}H_{c2}(T) =  \mu_{0}H_{c2}(0) \frac{1 - (T/T_\mathrm{C})^2}{1+(T/T_\mathrm{C})^2}
    \label{eqn 4:hc2}
\end{equation}
where $\mu_{0}H_{c2}(T)$ is temperature-dependent upper critical field.
The estimated upper critical field $\mu_{0}H_{c2}$(0) is  
7.14 T by extrapolating the GL fit. Furthermore, the Ginzburg-Landau coherence length ($\xi_{GL} (0) $)  is calculated using the following relation \cite{54,55}:
\begin{equation}
 \mu_{0}H_{c2}(0) = \frac{\phi_0} {2\pi\xi_\mathrm{GL}^2(0)}
    \label{eqn 5:GLeq}
\end{equation}
where $\phi_0$ is the magnetic flux quantum number ($h/2e$) and equals $2.07 \times 10^{-15}$ \text{T $m^2$}. By using  $\mu_{0}H_{c2}$(0), the value of $\xi_{GL}$(0) comes to be 6.79 nm. 
The superconducting state in type II superconductors are  completely destroyed above $\mu_{0}H_{c2}$(0). The suppression of superconductivity in type-II materials is governed by two different mechanisms: the orbital pair-breaking effect and the spin-paramagnetic pair-breaking effect. The orbital pair breaking mechanism arises when the kinectic energy of the Cooper pairs exceed the condensation energy while spin paramagnetic mechanism is due to the Zeeman splitting of the Copper pairs in the applied magnetic field. The orbital limiting field $\mu_{0}H_{c2}$(0) for BCS superconductors can be approximated by the WHH 
model \cite{12,54,55} and given as:   
\begin{equation}
    \mu_{0}H_{c2}^{\text{orbital}}(0) = -\alpha T_C \left. \frac{d\mu_{0}H_{c2}(T)}{dT}\right|_{T=T_C}
    \label{eqn 7:Hc2orbital}
\end{equation}
 In this equation, $\alpha$ is a dirty limit constant and equals 0.693. Here, we calculated the orbital pair breaking field  by using the slope $-\frac{dH_{c2}}{dT}$= -1.11 at $T=T_{C}$ and obtained as  $H_{c2}^{orbital}(0) =5.92$ T. The Pauli paramagnetic field can be calculated using $\mu_{0}H_{c2}^{P}(0)=1.85 T_{C}$ and equals to 14.25 T by taking $T_{C} =7.70K$ (considering $T_{C}^{mid}$). The relative strength of orbital and spin-paramagnetic effects are quantified using the Maki parameter $\alpha_M$ and given as: 
\cite{56} :
 \begin{equation}
     \alpha_M = \sqrt{2}\frac{H_{c2}^{orbital}(0)}{H_{c2}^{pauli}}
       \label{eqn 8:Maki}
 \end{equation}
 The obtained $\alpha_M$ is 0.59, indicating superconductivity is limited by orbital limiting field.
 To examine the contributions of orbital and pauli-paramagnetic effect, the experimental $\mu _0H_{c2} (T)$ is modelled using the WHH model \cite{54,19} and  is given by using following equation:
\begin{align} \ln\left(\frac{1}{t}\right) &= \left( \frac{1}{2} + \frac{i\lambda_{\mathrm{SO}}}{4\gamma} \right) \psi\left( \frac{1}{2} + \frac{\bar{h}+\lambda_{\mathrm{SO}}}{2} + \frac{i\gamma}{2t} \right) \nonumber\\ &\quad+ \left( \frac{1}{2} - \frac{i\lambda_{\mathrm{SO}}}{4\gamma} \right) \psi\left( \frac{1}{2} + \frac{\bar{h}+\lambda_{\mathrm{SO}}}{2} - \frac{i\gamma}{2t} \right) -\psi\left(\frac{1}{2}\right) \end{align}
where $t= T/T_{c}$ is the reduced temperature and $\psi(x)$ is the digamma function.
\begin{equation}
  \gamma= \Bigg[(\alpha_{M}\hbar)^2 - \bigg(\frac{1}{2}\lambda_{so} \bigg)^2\bigg]^{1/2} ,  
   \hbar = \frac{4H_{c2}}{\pi^2} \bigg (\frac{dH_{c2}}{dT}\bigg ) ^{-1}  
\end{equation}
where $\alpha_{M}$ is the Maki parameter and $\lambda_{so}$ is the spin-orbit interaction parameter. The WHH fit yields $\mu _0H_{c2} (0)=5.61 $ T without spin--orbit effect ($\lambda_{so} =0$ and $\mu _0H_{c2} (0)=6.43 $ T by considering spin--orbit effect ($\lambda_{so} =2.4$). The upper critical field derived from WHH considering the spin--orbit effect ($\lambda_{so} =2.4$) value is comparable to the value obtained using eq.\ref{eqn 7:Hc2orbital}.
 The magnetic susceptibility measurements were perforemed in field-cooled warming (FCW) and zero-field-cooled warming (ZFCW) modes over the temperature range of 1.8- 12 K, as shown in Figure \ref{Fig2:MT} (b).
The superconducting onset temperature $T_{c}^{\mathrm{onset}}$ obtained from the ZFCW measurement is found to be 7.51 K. The separation of ZFCW and FCW red magnetization curves indicates irreversibility due to strong vortex pinning. To determine the lower critical field ($\mu_{0}H_{c1}(0)$), isothermal magnetization measurements were performed over the temprature range 2K-5.5 K. The corresponding magnetization curves are presented in the inset of Figure~\ref{Fig3:Resistivity}(a). The deviation from Meissner state is taken as $\mu_{0}H_{c1}(T)$ at different temperatures. The estimated $\mu_{0}H_{c1}$ (0) is 0.023 T, using the Ginzburg-Landau approximation as:
\begin{equation}
 H_{c1}(T) = H_{c1}(0)\left[1-\left(\frac{T}{T_C}\right)^2\right]  
  \label{eqn 3:hc1}
\end{equation}
Also, the Ginzburg-Landau  penetration depth ($\lambda_{GL} (0)$) is evaluated using relationship between lower critical field $\mu_{0}H_{c1}(0)$ and coherence length $\xi_{GL}(0)$  and given as follows \cite{55,57} : 
\begin{equation}
   \mu_{0} H_{c1}(0) = \frac{\Phi_{0}}{4\pi\lambda_\mathrm{GL}^2(0)}\left(\ln\frac{\lambda_\mathrm{GL}(0)}{\xi_\mathrm{GL}(0)}+0.12\right)
      \label{eqn 6:hc1(0)}
\end{equation}
The calculated value of $\lambda_{GL} (0)$ from above eq. \ref{eqn 6:hc1(0)} is 151.5 nm, using $\mu_{0}H_{c1}$(0 ) = 0.023 T and $\xi_{GL}(0)$ =6.79 nm. In order to classify this material as a type I or type II superconductor, we calculated the Ginzburg-Landau parameter ($\kappa_{GL}$), which is defined as $\kappa_{GL}$ = $\frac{\lambda_{\mathrm{GL}}(0)}{\xi_{\mathrm{GL}}(0)}$. The estimated value of $\kappa_{GL}$ is 22.31, which is more than $\frac{1}{\sqrt{2}}$, confirming type II superconductivity. Also, the temperature dependence of the lower critical field was analyzed using the single-root Ginzburg--Landau expression, which is given as \cite{58} :
\begin{equation}
\mu_{0}H_{c1}(T)=
\frac{\phi_{0}}{4\pi}
\frac{\ln\!\left(1+\sqrt{2}\frac{\lambda_{\mathrm{GL}}(0)}{\xi_{\mathrm{GL}}(0)}\right)}
{\lambda_{\mathrm{GL}}^{2}(0)}
\left(1-\left(\frac{T}{T_C}\right)^2\right)
\label{eq:Hc1fit}
\end{equation}
where, $\lambda_{\mathrm{GL}}(0)$ is the Ginzburg--Landau penetration depth at zero temperature, $\xi_{\mathrm{GL}}(0)$ is the Ginzburg--Landau coherence length, and $T_C$ is the superconducting transition temperature. The experimental $\mu_0H_{c1}(T)$ data were fitted using eq.~(\ref{eq:Hc1fit}), with $\lambda_{\mathrm{GL}}(0)$ as free parameter, while $\xi_{\mathrm{GL}}(0)$ and $T_C$ were fixed to the values obtained from the upper critical field analysis and resistivity measurements, respectively, as shown in Figure~\ref{Fig3:Resistivity} (d). The best fit as shown in Figure~\ref{Fig3:Resistivity} (d) yields $\lambda_{\mathrm{GL}}(0)=158.49$ nm. This fitted value is in close agreement with the value obtained using eq. \ref{eqn 6:hc1(0)}.\\
Furthermore, heat capacity measurements are performed to elucidate the material's superconducting nature. The heat capacity data show a jump from the superconducting to the normal state at $T_{c}=7.6 K$, consistent with magnetization and resistivity data. Figure \ref{fig5:Specific heat} (a) shows the specific heat data plotted as $C/T$ versus $T^2$. The normal state data is fitted using the following expression:
\begin{equation}
C = \gamma_n T + \beta_1 T^3 + \beta_2 T^5
\label{eqn 10:Debyemodel}
\end{equation}
where $\gamma_n$ is the Sommerfeld coefficient related to the electronic contribution (normal state) to specific heat, and $\beta_1$ and $\beta_2$ corresponds to the phononic and anharmonic contribution to specific heat. The  $\gamma_n$, $\beta_1$ and $\beta_2$ are 4.22 $mJ mol^{-1}K^{-2}$, 0.1257 $mJ mol^{-1}K^{-4}$ and 0.00362 $\mu J mol^{-1}K^{-6}$ respectively. The obtained value of $\beta_1$ is used to calculate Debye temperature $\Theta_D$  according to the equation:
\begin{equation}
    \Theta_D^3  = \frac{12\pi^4RN}{5\beta_1} 
      \label{eqn 11:Thetad}
    \end{equation}
    where R= 8.314 $Jmol^{-1}K^{-1}$ is the universal gas constant and N is the total number of atoms in formula unit \cite{59}. The Debye temperature is 248.96 K, comparable to other HEAs and MEAs \cite{12}.
 The electronic density of state ($D(E_F)$)  at the Fermi level is calculated by using Sommerfeld coefficient ($\gamma_n$) as follows:
   \begin{equation}
    \gamma_n = \frac{\pi^2k_{B}^2}{3} D(E_F) 
      \label{eqn 13:DOS}
   \end{equation}
   where $k_B$ is the Boltzmann's constant ($1.38 \times 10 ^{-23} JK^{-1}$) and $\gamma_n$ is 4.22 $mJ mol^{-1}K^{-2}$\cite{18}.
   The computed density of state at fermi level using eq. \ref{eqn 13:DOS} is 1.79 states/eV f.u.. To analyze the \linebreak nature of the superconducting state and pairing mechanism, electronic specific heat capacity $(C_{el})$ is calculated by subtracting phononic contribution from measured total specific heat capacity and given as $C_{el}(T)= C_{tot}(T)-\beta_1 T^3 - \beta_2 T^5$. The electronic contribution $(C_{el})$ is well fitted with isotropic s-wave BCS model as shown in Figure \ref{fig5:Specific heat}(b) and  given as:
\begin{equation}
\frac{S}{\gamma_n T_C} =
-\frac{6}{\pi^2}
\left(
\frac{\Delta (0)}{k_{B}T_{C}}
\right)
\int_{0}^{\infty}
\left[
f \ln f + (1-f)\ln(1-f)
\right] dy
\label{eqn 14:BCSentropy}
\end{equation}

where f represents Fermi-Dirac distribution function and is given in terms of quasiparticle occupation index as $f(\xi) = \left[\exp\left(\frac{E(\xi)}{k_B T}\right) + 1\right]^{-1}$, quasiparticle energy $E(\xi) = \sqrt{\epsilon^2 + \Delta(t)^2}$,$y=\xi / \Delta(0)$, $t=T/T_C$ and  $ \Delta(t) = \Delta(0) \tanh\left[ 1.82 \left[ 1.018 \left( t^{-1} - 1 \right) \right]^{0.51} \right]$ is the defined as superconducting gap \cite{49}.
 Using eq.~(\ref{eqn 14:BCSentropy}), we have estimated the superconducting gap ratio
$\frac{\Delta(0)}{k_{B}T_{C}}=2.01$. The obtained value exceeds the weak coupling BCS limit of 1.76, indicating the strong coupling effects in the present alloy. The obtained value is larger than the weak-coupling BCS value of 1.76, indicating the presence of strong-coupling effects. To further examine the thermodynamic consistency of the superconducting state, the strong-coupling formalism is employed \cite{60}. The logarithmic average phonon frequency, $\omega_{\ln}$, is estimated using the experimentally determined normalized specific-heat jump, $\Delta C_{el}/\gamma_nT_C = 1.972$, through the relation
\begin{equation}
\frac{\Delta C_{el}}{\gamma_nT_C}
=
1.43\left[
1+53\left(\frac{T_C}{\omega_{\ln}}\right)^2
\ln\left(\frac{\omega_{\ln}}{3T_C}\right)
\right]
\label{eq:Carbotte_SH}
\end{equation}
which yields $\omega_{\ln}=115~\mathrm{K}$ using $T_C = 7.70K $ . Using   $\omega_{\ln}=115~\mathrm{K}$ and $T_C = 7.70K $, $\frac{\Delta(0)}{k_{B}T_{C}}$ is further evaluated as :
\begin{equation}
\frac{2\Delta(0)}{k_BT_C}
=
3.53\left[
1+12.5\left(\frac{T_C}{\omega_{\ln}}\right)^2
\ln\left(\frac{\omega_{\ln}}{2T_C}\right)
\right]
\label{eq:Carbotte_gap}
\end{equation}
resulting in $\frac{\Delta(0)}{k_BT_C}=1.96$. Further, the coupling strength of elctron-phonon interaction ($\lambda_{el-ph}$) was estimated using the Allen--Dynes equation \cite{60,61} and given as:
\begin{equation}
T_C=\frac{\omega_{\ln}}{1.2}
\exp\left[
-\frac{1.04(1+\lambda_{el-ph})}
{\lambda_{el-ph}-\mu^{*}\left(1+0.62\lambda_{el-ph}\right)}
\right],
\label{eq:AllenDynes}
\end{equation}
where $T_C = 7.70K $, $\omega_{\ln}$ is 115 K , and $\mu^{*}$ is the Coulomb pseudopotential and taken as 0.13 \cite{61}. The calculated value of the electron--phonon coupling constant $\lambda_{el-ph}$, using eq. \ref{eq:AllenDynes} is 1.07, indicating that the present system lies in the strong-coupling regime. 

\subsection*{Critical Current Density and Pinning Mechanism}
\begin{figure*}[!t]
   \begin{flushleft}
\includegraphics[width=1\linewidth,keepaspectratio]{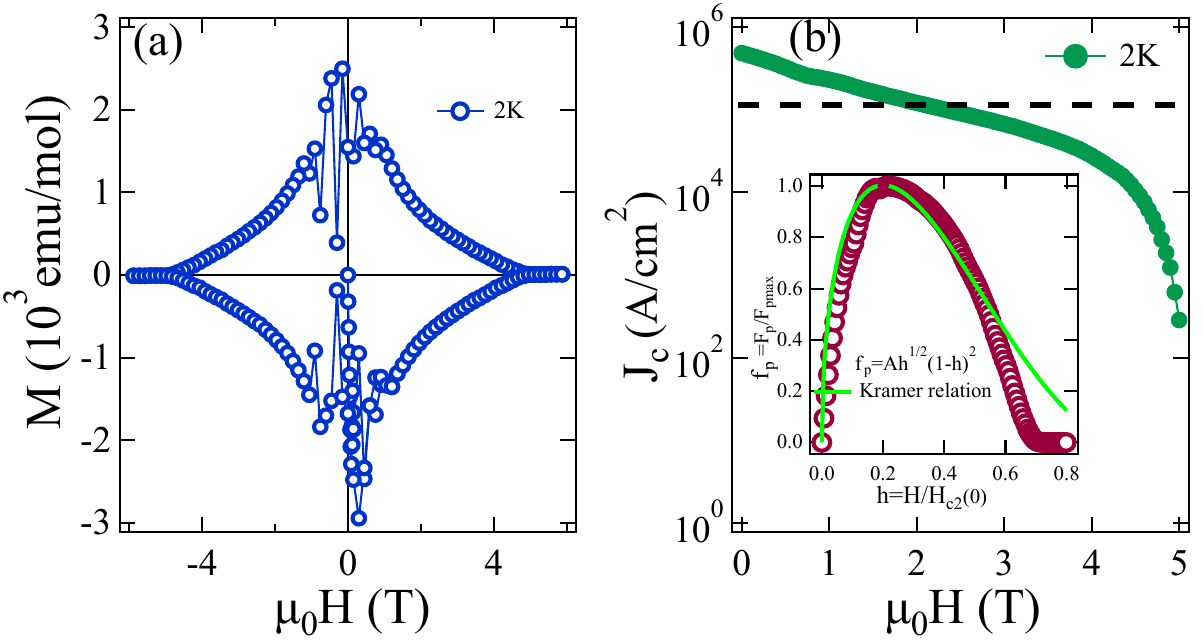}
    \end{flushleft}
    \vspace{-8 pt}
    \caption{(a) Isothermal magnetization hystersis loop (M-H) at T = 2K, used to calculate critical current density ($J_{c}$)  (b) Magnetic field dependent $J_{c}$, determined using Bean's model (eq. \ref{eqn9:Jc}). The inset shows normalized pinning force \(f_p = F_p / F_{p \text{max}}\) as a function of the normalized magnetic field \(h = H / H_{c2}(0)\) and scaled using the Dew-Hughes model to elucidate the dominant vortex pinning mechanism.}
    \label{fig4:MH}  
\end{figure*}

Figure \ref{fig4:MH} (a) shows the magnetic hysteresis loop (M-H) of \sample{} at 2K. The magnetic hysteresis curve exhibits type II superconductivity. The $J_{c}$ , is a key parameter for potential applications of superconductors and is calculated from the magnetization curve, using Bean's model  \cite{62} and given as :
 \begin{equation}
J_c =  \dfrac{2\Delta M}
{a\left(1-\dfrac{a}{3b}\right)}
\label{eqn9:Jc}
\end{equation}

where $\Delta M $ is the vertical magnetization width and a and b are the sample dimensions such that a<b. The sample dimensions are 1mm and 2mm, respectively, in our case. The variation of $J_{c}$ with field is shown in Figure \ref{fig4:MH} (b). The obtained value of $J_{c}$ is $0.48\times10^6 A/cm^{2}$. This value of $J_{c}$ exceeds the benchmark of $10^5 A/cm^{2}$, required for practical applications. Remarkably, the $J_{c}$ value surpasses the value obtained for other previously reported as-cast HEAs and MEAs. For instance, $\mathrm{Nb}_{0.4}\mathrm{Hf}_{0.2}\mathrm{Zr}_{0.2}\mathrm{Ti}_{0.2}$ shows $J_c$ of $1\times 10^{4} A/cm^2$ \cite{63}. In 3d-rich $\mathrm{Sc}_{0.2}\mathrm{Ti}_{0.2}\mathrm{V}_{0.2}\mathrm{Nb}_{0.2}\mathrm{Cu}_{0.2}$ shows $J_{c}$ = $0.18\times10^6 A/cm^{2}$ at 2 K \cite{18}. In equiatomic $\mathrm{Nb}_{0.25}\mathrm{Ta}_{0.25}\mathrm{Ti}_{0.25}\mathrm{Zr}_{0.25}$ critical current density is $\approx 0.13 \times10^{6} A/cm^2$ at temperatures ranging from 2 K to 4K \cite{19}. The eutectic $\mathrm{Nb}\mathrm{Sc}\mathrm{Ti}\mathrm{Zr}$ exhibits $J_c$ exceeding $10^{6} A/cm^2$ while, the BCC $\mathrm{Nb_{2/6}}\mathrm{Ta_{1/6}}\mathrm{Ti_{1/6}}\mathrm{Zr_{1/6}}\mathrm{Hf_{1/6}}$ and $(\mathrm{Nb\mathrm{Ta}_{0.7})(\mathrm{Ti}\mathrm{Zr}}\mathrm{Hf}_{0.5})$ alloys exhibit $J_c$ of the order of $10^{5} A/cm^2$ at 2 K \cite{20}. Remarkably, the present material highlights the highest critical current density value observed in any reported pure-phase as-prepared HEAs and MEAs.
\\
 In order to elucidate the flux pinning mechanism, the normalized pinning force  \(f_p = F_p / F_{p \text{max}}\) is plotted as a function of the reduced magnetic field \(h = H / H_{irr}(T)\) and shown in the Figure \ref{fig4:MH} (c). Using $J_c$ value, the pinning force $F_p$ is given as $F_p \propto\mu_0 H J_c$. The scaling behavior of $f_p$ is described by using the Dew-Hughes model as $f_{p}\propto h^{p}(1-h)^{q}$ \cite{64}, where p and q are fitting parameters to provide information about the flux pinning mechanism. The experimental data is fitted with the characteristic exponents $p=0.5$, $q=2$ for surface pinning and $p=1$, $q=2$ for point pinning. The fitting results indicate that the flux pinning behavior in the present alloy is mainly governed by the point pinning contribution. The severe lattice distortion discussed in the latter part of the manuscript could enhance the pinning effect at grain boundaries and therefore cause high $J_c$ in this sample.
 \subsection*{Strongly Correlated Behaviour}
\begin{figure*}[!t]
\includegraphics[width=1\linewidth,keepaspectratio]{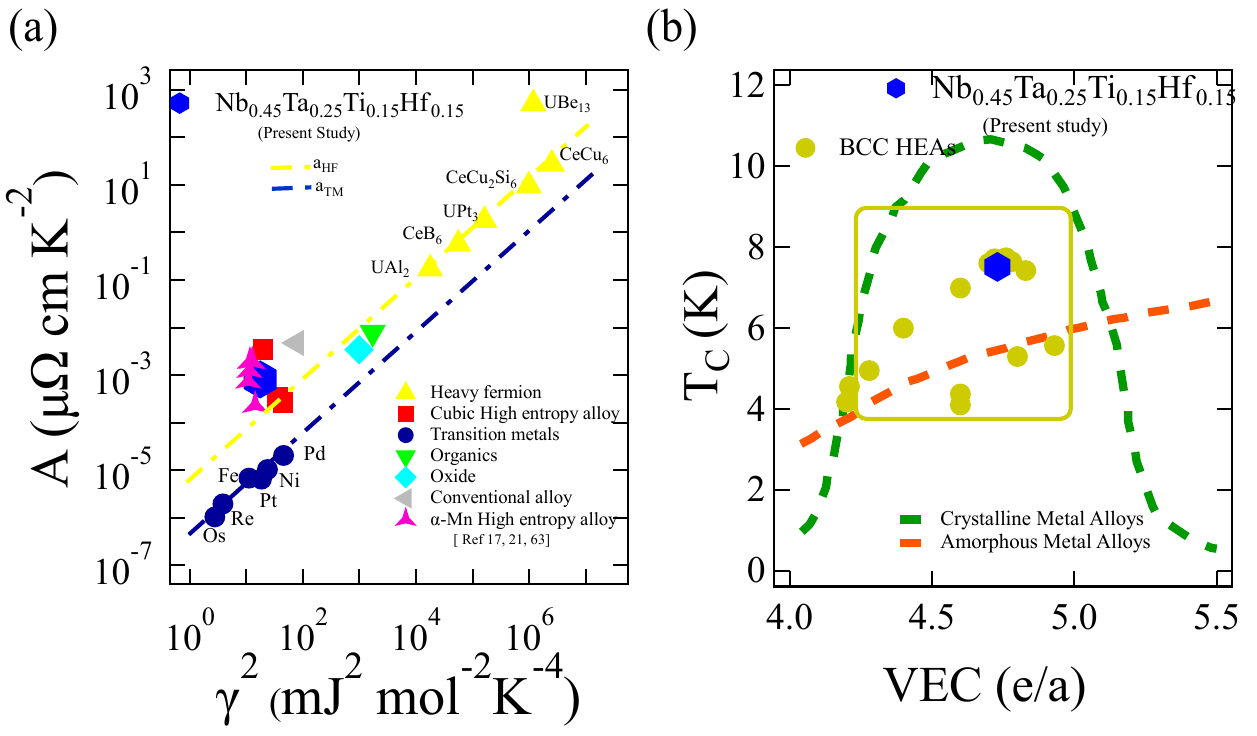}
    \caption{(a) Kadowaki-Woods ratio (KWR) plot of A versus $\gamma^{2}$ for \sample, compared with other compounds, including HEAs, heavy fermion system and other materials \cite{21,57,65,66}. (b) Correlation between Valence Electron Count (VEC) and transition temperature $T_C$ for various crystalline metal alloys, amorphous metal alloys and BCC HEAs \cite{67,68}.} 
    \label{fig:KWR and VEC}  
\end{figure*}

The Kadowaki--Woods ratio (KWR) is calculated to investigate the correlation between the electron--electron scattering coefficient (A) and quasiparticle mass renormalization factor in terms of $\gamma_n$, thereby probing electronic correlations. Figure~\ref{fig:KWR and VEC}(a) represents the KWR plot of $A$ vs.\ $\gamma_{n}^{2}$ for various materials, where $A$ is the coefficient of $T^2$ obtained from the low-temperature resistivity ($\rho = \rho_{0} + AT^2$) and $\gamma_{n}$ is obtained from the electronic specific heat. The KWR is defined as $A/\gamma^2_n$. For transition metals the characteristics value of KWR is $a_{\mathrm{TM}} = 0.4~\mu\Omega\,\mathrm{cm\,mol^{2}\,K^{2}\,J^{-2}}$, whereas in heavy fermion system it is significantly enhanced and reaching $a_{\mathrm{HF}} = 10~\mu\Omega\,\mathrm{cm\,mol^{2}\,K^{2}\,J^{-2}}$   \cite{65,66}.
For Nb$_{0.45}$Ta$_{0.25}$Ti$ _{0.15}$ \linebreak Hf$_{0.15}$, the KWR is $45.48 \mu\Omega\,\mathrm{cm\,mol^{2}\,K^{2}\,J^{-2}}$, placing the sample near heavy-fermion line \cite{21,57}. This large ratio of $A$/\ $\gamma_{n}^{2}$ indicates pronounced electron--electron correlations and classifies Nb$_{0.45}$Ta$_{0.25}$  \linebreak Ti$ _{0.15}$Hf$_{0.15}$ as a strongly correlated material. The large KWR value is likely to arise from enhanced disorder, and electron-phonon interactions \cite{57}. Strongly correlated behavior is uncommon in alloy superconductors, and most of the time, they highlight anomalous superconducting behavior \cite{69}. Therefore, in the current MEA, the emergence of strongly correlated behavior opens a promising opportunity to explore the interplay among superconductivity, electronic correlations, and extreme disorder \cite{21,57}.
 
 Figure \ref{fig:KWR and VEC} (b) shows the variation of transition temperature $T_C$ with valence electron Count (VEC) for various crystalline metal alloys, amorphous alloys, and  BCC HEAs. According to the Matthias rule, BCC HEAs exhibit the highest $T_C$ within the VEC range of $4.67$--$4.70$. The present sample is consistent with the VEC range reported for other HEAs \cite{67,68}.
\begin{figure*}
    \includegraphics[width=1\linewidth]{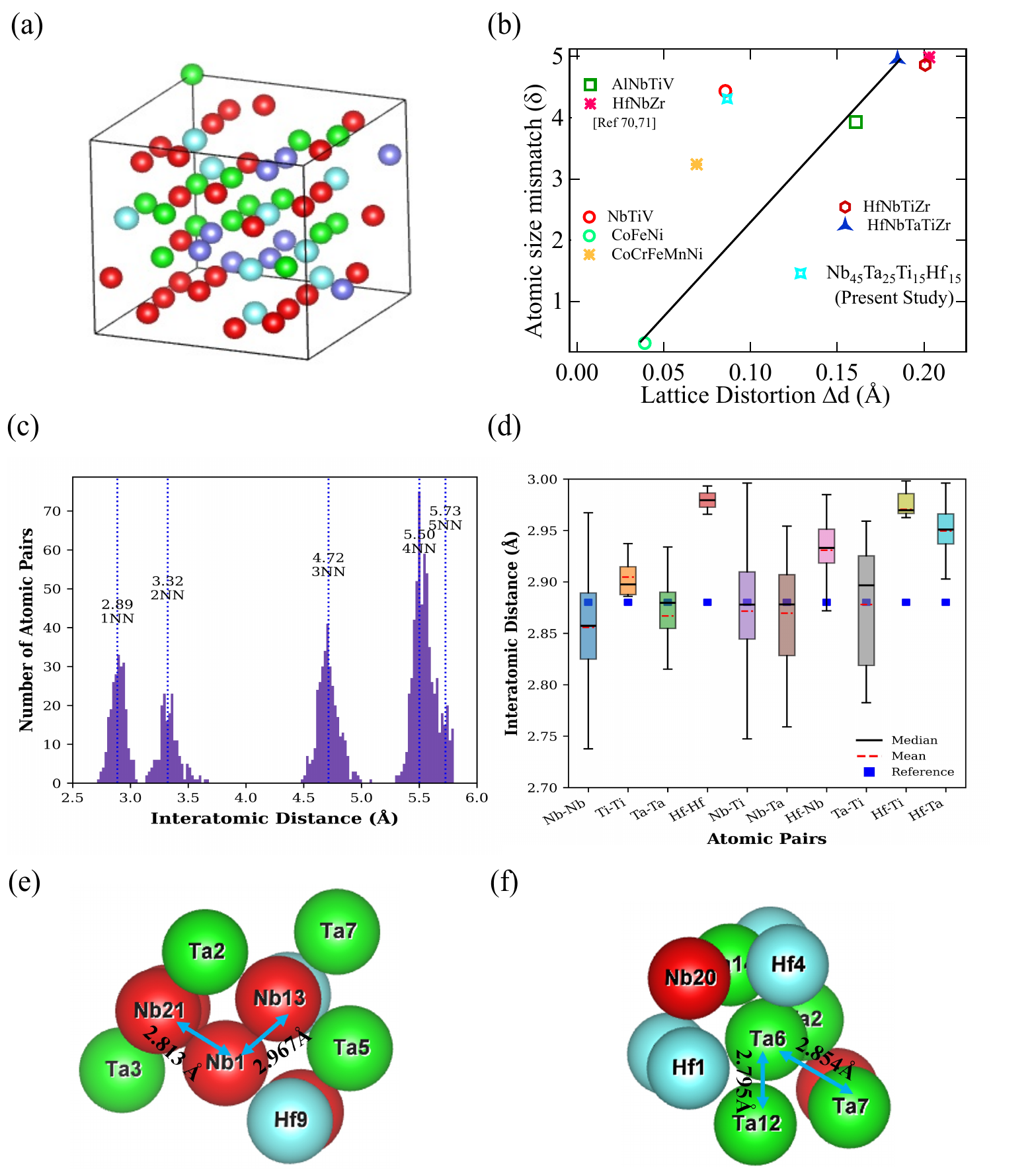}
    \caption{(a) Schematic structural representation of a 60-atom Special Quasi Random structure (SQS) supercell  (b) The relationship between atomic size mismatch $\delta$, estimated from elemental atomic radii (Eq.~15), and the \textit{first principles} calculated mean atomic displacement $\Delta d$ (Eq.~16) is illustrated \cite{70,71} in . (c) Interatomic distance distribution from 1NN to 5NN. Dashed lines mark distinct peaks for the ideal BCC structure, while the relaxed SQS structure shows partial filling of the gaps. (d) The distribution of first nearest-neighbour (1NN) interatomic distances for various atomic pairs in the \sample is depicted using box plots. The mean interatomic distance, median bond length and refrence bond length are represented by a red dashed line, a solid green line and a blue solid rectangle, respectively. (e - f) The shortest and longest first nearest neighbor bond lengths for Nb-Nb and Ta-Ta pairs in SQS supercell.}
     \label{fig8:LLD_P6}  
\end{figure*}
\begin{figure*}
    \centering
    \includegraphics[width= 1 \linewidth]{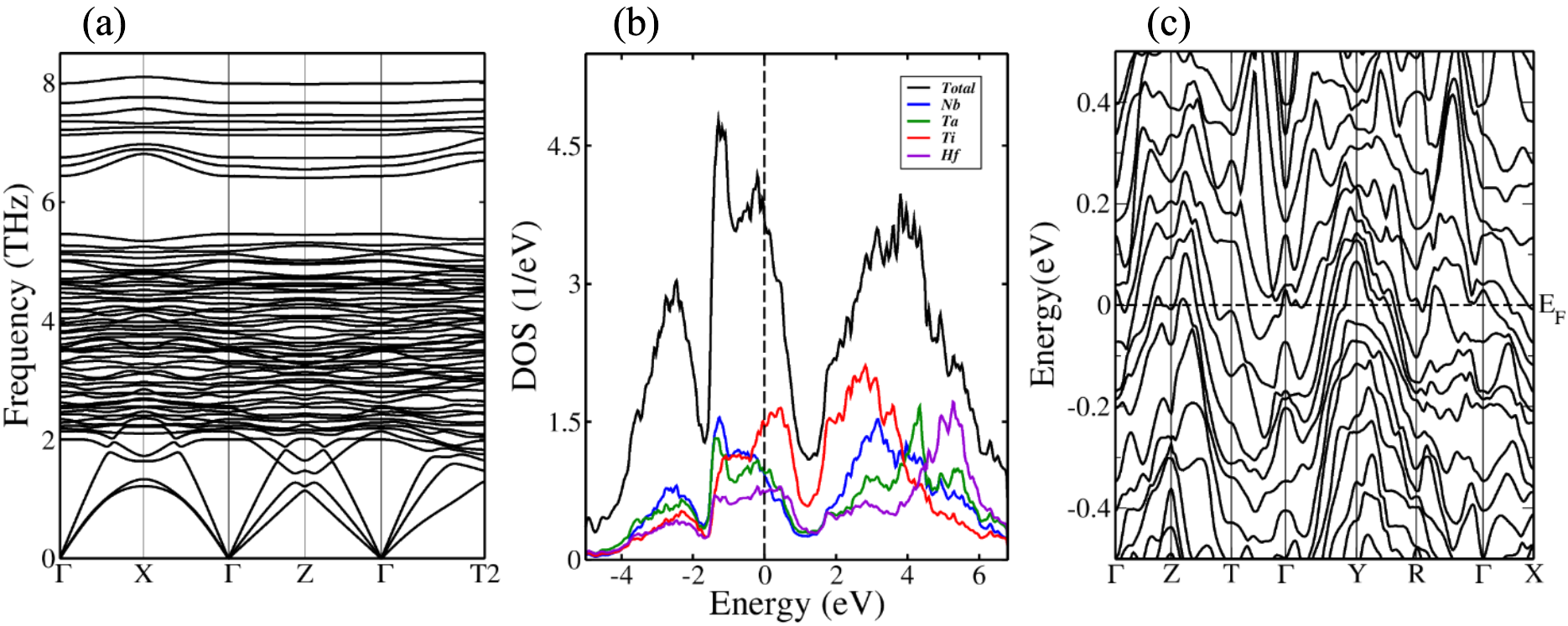}
    \caption{(a) Calculated phonon dispersion of \sample along selected high-symmetry directions in the Brillouin
zone, (b) total density of states (DOS) and atom-resolved partial DOS projected onto the d-orbitals of each constituent transition-metal element (Nb, Ta, Ti, and Hf), (c) band-dispersion along high-symmetry directions.}
    \label{phonon&dos&electronic_band}
\end{figure*}
\subsection*{Lattice Distortion}
HEAs and MEAs  exhibit severe lattice distortion due to the presence of multicomponent elements having different sizes \cite{72,73}. In order to quantify the local lattice distortion in BCC \sample, interatomic distances are calculated using 60 atoms SQS as shown in Figure \ref{fig8:LLD_P6} (c).  The broadness of the interatomic distances plot indicates that atoms deviate significantly from their mean positions, causing pronounced lattice distortion. This peak broadness is attributed to size mismatch of constituent elements as $r_{Ti} \leq r_{Nb} \leq r_ {Ta} \leq r_{Hf}$. The average interatomic distance for the first five NN pairs is 2.89, 3.32, 4.72, 5.50, and 5.73 \AA, respectively. However, the peak position is still centered around their ideal interatomic distances, predicting crystalline behaviour is preserved over long-range. Figure \ref{fig8:LLD_P6} (a) shows a 60-atom supercell of the BCC structure using SQS. The atomic size mismatch is commonly used to quantify the effects of local lattice distortion (LLD) in HEAs and MEAs, using the equation:
\begin{equation}
\delta = \sqrt{\sum_{i=1}^{n} c_i \left(1 - \frac{r_i}{\bar{r}} \right)^2}
\label{eqn 15:delta}
\end{equation}
where $\bar{r}$ being the average atomic radius of the constituent elements and given as  $\sum_{i=1}^{n} c_i r_i$, where $c_i$ and $r_i$ are atomic fraction and atomic radius of individual elements, respectively. This approach yields somewhat ambiguous results, as atomic radii depend on factors such as the nature of bonding, crystal structure, and external conditions \cite{70}. Hence, for accurate measurement of local lattice distortion with atomic radii measured under different conditions, we measured the mean atomic displacement as follows:
\begin{equation}
\Delta d =  \frac{1}{N}\sqrt{ \sum_{i=1}^{N} \left[ (x_i - x_i')^2 + (y_i - y_i')^2 + (z_i - z_i')^2 \right] } 
  \label{eqn 16:LLD}
\end{equation}
where $(x_{i}, y_{i}, z_{i})$ and $(x_{i}', y_{i}', z_{i}')$ are the reduced coordinates of the $i$ th atom before and after the structural relaxation. The obtained value of mean atomic displacement ($\Delta d$) is 0.086 \AA, which indicates the displcement of atoms from their ideal positions. This result predicts that $\delta$, as a radius-dependent parameter, can overestimate the lattice distortion, while $\Delta d$ gives a more precise intrinsic structural response to chemical disorder. Figure \ref{fig8:LLD_P6} (b) shows a linear dependence of  the $\delta$ and  $\Delta d$. Typically, a larger atomic size mismatch ($\delta$) is associated with a greater local lattice distortion ($\Delta d$). Nevertheless, in some 3d HEAs and MEAs, $\delta$ tends to slightly overestimate the local lattice distortion (LLD) when it is calculated using the atomic radii $r_i$ of the individual elements  \cite{71}.\\
To get deeper insight of microscopic origin of LLD, a pair-wise analysis of fist nearest neighbour (1NN) shell was carried out and presented in Figure \ref{fig8:LLD_P6} (d). The BCC alloy{} Nb$_{0.45}$Ti$_{0.15}$ Ta$_{0.25}$Hf$_{0.15}$ composed of both homoatomic as well as heteroatomic pairs of atoms. The homoatomic pairs include Nb-Nb, Ta-Ta, Ti-Ti, Hf-Hf while heteroatomic pairs composed of Nb-Ti, Nb-Ta, Nb-Hf, Ti-Ta, Ti-Hf and Ta-Hf. The computed results yield comparable bond lengths for MEAs and HEAs. The mean and median corresponding to each pair are shown in Figure \ref{fig8:LLD_P6} (d). The pair-wise LLD can be evaluated using the formula:\\
 \begin{equation}
 \mathrm{{Lattice \ distortion = \frac{d_{i}-d_{r}}{d_{r}} } }
 \label{eq. lattice distortion}
 \end{equation}
where $d_{i}$ corresponds to the interatomic distance of the $i$ th atom in relaxed SQS supercell and $d_r$ is the reference distance in unrelaxed structure. Using eq. \ref{eq. lattice distortion}, the maximum distortion is shown by the Hf--Hf pair (3.45\%) and the Ti--Ti pair (0.85\%), respectively, reflecting the larger atomic radius and correspondingly stronger local lattice strain of Hf compared with Ti in the BCC environment. The higher value for Hf--Hf arises from the pronounced size-mismatch, driven expansion when two Hf atoms occupy nearest-neighbor sites, while the smaller Ti--Ti distortion indicates that Ti induces only a moderate local strain due to its relatively smaller size mismatch with the BCC lattice. A similar trend was observed in $(TaNb)_{0.67}(HfZrTi)_{0.33}$ \cite{74}. Among heteroatomic pairs, the largest distortion is exhibited by Hf-Ti and Hf-Ta, which are respectively 3.13\% and 2.40\%. The atomic pair is compressed when the median bond length is shorter than corresponding reference bond length of the pair e.g., Nb-Nb pair. The heteroatomic pairs such as Nb-Ti and Nb-Ta shows less deviations from their neighbouring atoms, indicating a balanced atomic environment. The heteroatomic pairs, such as Nb with Ti and Ta, appear to be balanced with their neighboring atoms. The elongation or compression of bond lengths is governed by the local environment of a atomic pair, where certain neighboring atoms induce compression while others cause elongation of the bonds. Also, changes in bond length are related to chemical interactions among multiprincipal elements in MEAs. Further, the shortest and longest first nearest neighbor bond lengths for Nb-Nb and Ta-Ta pairs in SQS supercell is shown in Figure \ref{fig8:LLD_P6} (e-f).    
\begin{table}[h!]
\caption{Experimentally determined superconducting parameters of \sample}
\label{superconducting properties}
\begin{center}
\begin{tabular*}{1.0\columnwidth}{l@{\extracolsep{\fill}}lll}\hline\hline
Properties& unit& \sample\\
\hline
                                 
  T$^{onset}_{C}$ ($\rho$ - T)  & K&  7.8 (1)&\\ 
  T$^{mid}_{C}$ ($\rho$ - T)  & K&  7.7 (1)&\\ 
$\mu_0$H$_{c1}$(0)& T& 0.0232 (1) & \\
$\mu_0$H$_{c2}$(0)& T& $ 7.14 (9) $& \\         $\mu_0$H$_{c2}$(0) (WHH) & T& $ 7.14 (9) $& \\              
$\mu_0H_{c2}^{P}(0)$& T& 14.33& \\
$\xi _{GL}(0)$& \text{\AA}& 67.9 (4) & \\
$\lambda _{GL}(0)$& \text{\AA}& 1515 (3) & \\
$\kappa _{GL}(0)$& &22.31  &\\
$\gamma$ & mJ mol$^{-1}$ K$^{-2}$ & $4.22  (6)$ \\
$\beta_1$ & mJ mol$^{-1}$ K$^{-4}$ & $0.1257(6)$ \\
$\beta_2$ & mJ mol$^{-1}$ K$^{-4}$ & $0.0036 (62)$ \\
$J_C$& A/cm$^2$&  $0.48 \times10^6$& \\
$\theta_{D}$& K& 248.96 (4) &\\
$KWR$ & $\mu\Omega\,\mathrm{cm}\,\mathrm{mol}^{2}\,\mathrm{K}^{2}\,\mathrm{J}^{-2}$ & 45.48(12) & \\
$D(E_F)$& states/ev f.u& 1.79 (3) &\\
 $\omega_{ln}$  & K&  115 (6) &\\
 $\lambda_{el-ph}$& & 1.07 (3) &\\
$\Delta_{0}/k_{\mathrm{B}}T_C$ & & 2.01 & \\
$\Delta C_{el}/\gamma_nT_C$ & & 1.98 & \\
\hline\hline
\end{tabular*}
\par\medskip\footnotesize
\end{center}
\end{table}
\subsection*{First-Principles Studies of Structural Stability and Electronic Properties}
To gain microscopic insight into the experimentally observed structural stability and metallic normal-state behavior of \sample, we performed first-principles calculations within the generalized gradient approximation (GGA). Since the alloy possesses intrinsic substitutional disorder arising from the random occupation of lattice sites by multiple constituent elements, the chemical disorder was explicitly modeled using the special quasirandom structure (SQS) approach. The calculated lattice dynamical and electronic properties are summarized in Fig.~\ref{phonon&dos&electronic_band}. The calculated phonon dispersion [Fig.~\ref{phonon&dos&electronic_band}(a)] exhibits no imaginary phonon frequencies throughout the Brillouin zone, confirming the dynamical stability of the chemically disordered body-centered cubic phase. The optimized equilibrium lattice parameter is also in excellent agreement with the experimentally determined value, demonstrating that the SQS model faithfully captures the disordered crystal structure realized experimentally. The phonon spectrum extends up to approximately 8 THz, with acoustic modes below $\sim$2 THz followed by relatively flat optical branches at higher frequencies. Such broad phonon features originate from the coexistence of constituent atoms with different masses and local bonding environments, reflecting the intrinsic chemical disorder characteristic of entropy-stabilized alloys. These results establish that the chemically disordered BCC phase is both structurally and dynamically stable, consistent with the experimentally observed single-phase crystal structure.
 \par The corresponding electronic structure is shown in Fig.~\ref{phonon&dos&electronic_band}(b,c). The total and atom-resolved density of states reveal a finite electronic density at the Fermi level, confirming the metallic ground state of \sample. This is fully consistent with the experimentally observed metallic resistivity above the superconducting transition. The electronic states near $E_F$ are dominated by transition-metal $d$ orbitals, with Ti and Ta contributing the largest spectral weight, while Nb and Hf provide broader contributions over a wider energy range through strong orbital hybridization. The absence of a pseudogap or electronic instability near $E_F$ indicates that the pronounced chemical disorder does not disrupt the itinerant electronic states required for superconductivity. The calculated band structure further elucidates the low-energy electronic states responsible for the metallic behavior. Multiple dispersive bands cross the Fermi level along the high-symmetry directions, giving rise to several electron and hole pockets and indicating an intrinsically multiband electronic structure. Rather than arising from an individual atomic species, these low-energy bands originate from the hybridized $d$-orbitals of Nb, Ta, Ti, and Hf, reflecting the cooperative electronic nature of the chemically disordered alloy. Such multiband electronic states naturally account for the finite density of states at the Fermi level and provide the microscopic electronic framework for the experimentally observed metallic transport and superconductivity.
\par 
Thus, our first-principles calculations establish a coherent microscopic picture of \sample. The absence of unstable phonon modes confirms the stability of the chemically disordered BCC phase, while the multiband $d$-electron metallic state explains the robust normal-state conductivity that precedes superconductivity. Together with the experimentally observed strong electron-phonon coupling and superconducting transition, these calculations demonstrate that superconductivity in \sample\ emerges from a dynamically stable, chemically disordered metallic phase whose low-energy electronic structure is governed by strongly hybridized transition-metal $d$ states.
\section{Conclusion}  
In summary, we have carried out a comprehensive experimental and first-principles investigation of the structural, superconducting, electronic, and local lattice-distortion properties of the mechanically robust medium-entropy alloy \sample. Structural characterization confirms the formation of a homogeneous single-phase body-centered cubic structure with a nearly stoichiometric elemental distribution. Low-temperature magnetization, transport, and specific-heat measurements establish bulk type-II superconductivity with a transition temperature of $T_c^{\mathrm{onset}}\approx7.8$ K, an upper critical field of $\mu_0H_{c2}(0)=7.14\pm0.08$ T, and a strongly coupled isotropic $s$-wave superconducting state. Remarkably, \sample\ exhibits a critical current density of $0.48\times10^{6}$~A/cm$^2$, the highest reported among as-cast medium- and high-entropy superconducting alloys, highlighting its exceptional capability for sustaining large supercurrents. 
Photoemission spectroscopy confirms the metallic electronic character with finite spectral weight at the Fermi level, while GGA-based first-principles calculations establish the dynamical stability of the chemically disordered BCC phase and reveal a multiband metallic electronic structure dominated by hybridized Ti and Ta $d$ states near the Fermi level. The SQS calculations further quantify substantial disorder-induced local lattice distortion ($\Delta d=0.086$~\AA), providing a microscopic picture of the intrinsic structural disorder that is likely responsible for the enhanced vortex-pinning landscape and the observed high critical current density.
\par
Beyond its outstanding superconducting performance, the unusually large Kadowaki-Woods ratio places \sample\ close to the heavy-fermion regime, suggesting that significant electronic correlations coexist with strong electron-phonon coupling despite the presence of extreme chemical disorder. The concurrent presence of high critical current density, strong vortex pinning, enhanced electronic correlations, and exceptional mechanical robustness positions this material as a highly promising candidate for next-generation superconductors for operation under extreme conditions. 

\section*{Acknowledgments}
S. M. acknowledges financial support from the ANRF, Government of India, for the Advanced Research Grant (ANRF/ARG/2025/004870/PS). S. M. also acknowledges the financial support from the SPARC, Ministry of Education, Govt. of India (Project no. P4139). S.K.P. acknowledges support from ANRF (previously SERB), Government of India, for the core research grant (CRG/2023/003063)


\begin{thebibliography}{References}
\bibitem{1} C. Yao and Y. Ma, \textit{iScience}, 2021, \textbf{24}, 102541.
\bibitem{2} C. Eom, M. Lee, J. Choi, L. Belenky, X. Song, L. Cooley, M. Naus, S. Patnaik, J. Jiang, M. Rikel \textit{et al.}, \textit{Nature}, 2001, \textbf{411}, 558--560.

\bibitem{3} D. Larbalestier, A. Gurevich, D. M. Feldmann, A. Polyanskii, \textit{Nature}, 2001, \textbf{414}, 368--377.

\bibitem{4} B. Cantor, \textit{Entropy}, 2014, \textbf{16}, 4749--4768.

\bibitem{5} M.-H. Tsai and J.-W. Yeh, \textit{Materials Research Letters}, 2014, \textbf{2}, 107--123.

\bibitem{6} K. M. Youssef, A. J. Zaddach, C. Niu, D. L. Irving and C. C. Koch, \textit{Materials Research Letters}, 2015, \textbf{3}, 95--99.

\bibitem{7} E. P. George, D. Raabe and R. O. Ritchie, \textit{Nature Reviews Materials}, 2019, \textbf{4}, 515--534.

\bibitem{8} Y. Yuan, Y. Wu, Z. Yang, X. Liang, Z. Lei, H. Huang, H. Wang, X. Liu, K. An and W. Wu \textit{et al.}, \textit{Materials Research Letters}, 2019, \textbf{7}, 225--231.

\bibitem{9} S. Vrtnik, P. Koželj, A. Meden, S. Maiti, W. Steurer, M. Feuerbacher and J. Dolinšek, \textit{Journal of Alloys and Compounds}, 2017, \textbf{695}, 3530--3540.

\bibitem{10} J.-W. Yeh, S.-K. Chen, S.-J. Lin, J.-Y. Gan, T.-S. Chin, T. T. Shun, C.-H. Tsau and S.-Y. Chang, \textit{Advanced Engineering Materials}, 2004, \textbf{6}, 299--303.

\bibitem{11} L. Zeng, L. Li, K. Li, R. Chen and H. Luo, \textit{NPG Asia Materials}, 2024, \textbf{16}, 60.

\bibitem{12} P. Sobota, R. Topolnicki, T. Ossowski, T. Pikula, A. Pikul and R. Idczak, \textit{Physical Review B}, 2022, \textbf{106}, 184512.

\bibitem{13} F. Tian, L. Delczeg, N. Chen, L. K. Varga, J. Shen and L. Vitos, \textit{Physical Review B}, 2013, \textbf{88}, 085128.
\bibitem{14} D. Li, P. K. Liaw, L. Xie, Y. Zhang and W. Wang, \textit{Journal of Materials Science \& Technology}, 2024, \textbf{186}, 219--230.

\bibitem{15} O. El-Atwani, N. Li, M. Li, A. Devaraj, J. K. S. Baldwin, M. M. Schneider, D. Sobieraj, J. S. Wróbel, D. Nguyen Manh and S. A. Maloy \textit{et al.}, \textit{Science Advances}, 2019, \textbf{5}, eaav2002.

\bibitem{16} P. Koželj, S. Vrtnik, A. Jelen, S. Jazbec, Z. Jagličić, S. Maiti, M. Feuerbacher, W. Steurer and J. Dolinšek, \textit{Physical Review Letters}, 2014, \textbf{113}, 107001.

\bibitem{17} R. Hidayati, J. H. Kim, S.-G. Jung, K.-S. Cho, J. H. Yun and J.-S. Rhyee, \textit{Acta Materialia}, 2023, \textbf{261}, 119420.

\bibitem{18} S. Jangid, P. Meena, R. Kushwaha, S. Srivastava, P. Manna, S. Sharma, P. Mishra and R. Singh, \textit{Physical Review Materials}, 2025, \textbf{9}, 094804.

\bibitem{19} N. Sharma, K. Kargeti, N. Sharma, P. Chourasia, B. Vignolle, O. Toulemonde, T. Chakraborty, S. Panda and S. Marik, \textit{Physical Review B}, 2025, \textbf{112}, 224515.

\bibitem{20} J. Kitagawa, Y. Mizuguchi and T. Nishizaki, \textit{The European Physical Journal B}, 2025, \textbf{98}, 73.

\bibitem{21} G. Kim, M.-H. Lee, J. H. Yun, P. Rawat, S.-G. Jung, W. Choi, T.-S. You, S. J. Kim and J.-S. Rhyee, \textit{Acta Materialia}, 2020, \textbf{186}, 250--256.

\bibitem{22} T. Seki, H. Arima, Y. Kawasaki, T. Nishizaki, Y. Mizuguchi and J. Kitagawa, \textit{Journal of Superconductivity and Novel Magnetism}, 2024, \textbf{37}, 1059--1068.

\bibitem{23} S.-G. Jung, Y. Han, J. H. Kim, R. Hidayati, J.-S. Rhyee, J. M. Lee, W. N. Kang, W. S. Choi, H.-R. Jeon and J. Suk \textit{et al.}, \textit{Nature Communications}, 2022, \textbf{13}, 3373.

\bibitem{24} P. Sobota, B. Rusin, D. Gnida, R. Topolnicki, T. Ossowski, W. Nowak, A. Pikul and R. Idczak, \textit{Acta Materialia}, 2025, \textbf{285}, 120666.

\bibitem{25} D. H. Cook, P. Kumar, M. I. Payne, C. H. Belcher, P. Borges, W. Wang, F. Walsh, Z. Li, A. Devaraj and M. Zhang \textit{et al.}, \textit{Science}, 2024, \textbf{384}, 178--184.

\bibitem{26} P. Hohenberg and W. Kohn, \textit{Physical Review}, 1964, \textbf{136}, B864.

\bibitem{27} W. Kohn and L. J. Sham, \textit{Physical Review}, 1965, \textbf{140}, A1133.

\bibitem{28} A. Zunger, S.-H. Wei, L. G. Ferreira and J. E. Bernard, \textit{Physical Review Letters}, 1990, \textbf{65}, 353--356.

\bibitem{29} A. van de Walle, M. Asta and G. Ceder, \textit{Calphad}, 2002, \textbf{26}, 539--553.

\bibitem{30} A. van de Walle, P. Tiwary, M. de Jong, D. Olmsted, M. Asta, A. Dick, D. Shin, Y. Wang, L.-Q. Chen and Z.-K. Liu, \textit{Calphad}, 2013, \textbf{42}, 13--18.

\bibitem{31} A. van de Walle, \textit{Calphad}, 2009, \textbf{33}, 266--278.

\bibitem{32} A. van de Walle, M. Asta and G. Ceder, \textit{Calphad}, 2002, \textbf{26}, 539--553.

\bibitem{33} P. E. Blöchl, \textit{Physical Review B}, 1994, \textbf{50}, 17953--17979.

\bibitem{34} G. Kresse and D. Joubert, \textit{Physical Review B}, 1999, \textbf{59}, 1758--1775.
\bibitem{35} G. Kresse and J. Hafner, \textit{Physical Review B}, 1993, \textbf{47}, 558--561.

\bibitem{36} G. Kresse and J. Furthmüller, \textit{Computational Materials Science}, 1996, \textbf{6}, 15--50.

\bibitem{37} G. Kresse and J. Furthmüller, \textit{Physical Review B}, 1996, \textbf{54}, 11169--11186.

\bibitem{38} J. P. Perdew, K. Burke and M. Ernzerhof, \textit{Physical Review Letters}, 1996, \textbf{77}, 3865--3868.

\bibitem{39} J. P. Perdew, A. Ruzsinszky, G. I. Csonka, O. A. Vydrov, G. E. Scuseria, L. A. Constantin, X. Zhou and K. Burke, \textit{Physical Review Letters}, 2008, \textbf{100}, 136406.

\bibitem{40} H. J. Monkhorst and J. D. Pack, \textit{Physical Review B}, 1976, \textbf{13}, 5188--5192.

\bibitem{41} S. Baroni, S. de Gironcoli, A. Dal Corso and P. Giannozzi, \textit{Reviews of Modern Physics}, 2001, \textbf{73}, 515.

\bibitem{42} X. Gonze and C. Lee, \textit{Physical Review B}, 1997, \textbf{55}, 10355--10368.

\bibitem{43} A. Togo and I. Tanaka, \textit{Scripta Materialia}, 2015, \textbf{108}, 1--5.

\bibitem{44} M. Born and K. Huang, \textit{Dynamical Theory of Crystal Lattices}, Oxford University Press, 1996.

\bibitem{45} Q. Hu, S. Guo, J. Wang, Y. Yan, S. Chen, D. Lu, K. Liu, J. Zou and X. Zeng, \textit{Scientific Reports}, 2017, \textbf{7}, 39917.

\bibitem{46} S. Doniach and M. Šunjić, \textit{Journal of Physics C: Solid State Physics}, 1970, \textbf{3}, 285--291.

\bibitem{47} N. Bhatt, D. Sharma, A. Ali, K. Motla, S. Jangid, R. P. Singh and R. S. Singh, \textit{Physical Review Materials}, 2025, \textbf{9}, L092001.

\bibitem{48} J. E. Whitten, \textit{Applied Surface Science Advances}, 2023, \textbf{13}, 100384.

\bibitem{49} S. Jangid, P. Meena, R. Kushwaha, S. Srivastava, P. Manna, P. Mishra, S. Sharma and R. Singh, \textit{Applied Physics Letters}, 2024, \textbf{124}, 192602.

\bibitem{50} G. Grimvall, \textit{Physica Scripta}, 1976, \textbf{14}, 63.

\bibitem{51} Z. Xie, M. Yang, Z. G. Cheng, T. Ying, J.-G. Guo and X. Chen, \textit{Journal of the Physical Society of Japan}, 2023, \textbf{92}, 054702.

\bibitem{52} M. Krishnan, J. J. Meng, B.-Z. Li, Y. Ma, C. Wang, D. Bhoi, Y. Uwatoko, Q. Jing and B. Liu, \textit{Journal of Physics and Chemistry of Solids}, 2025, \textbf{202}, 112630.

\bibitem{53} N. Sharma, N. Sharma, T. Chakraborty and S. Marik, \textit{Materials Today Communications}, 2025, \textbf{44}, 111968.

\bibitem{54} S. Chakrabortty and N. Mohapatra, \textit{Physical Review B}, 2025, \textbf{111}, 094511.

\bibitem{55} P. Sobota, R. Topolnicki, T. Ossowski, T. Pikula, D. Gnida, R. Idczak and A. Pikul, \textit{Scientific Reports}, 2023, \textbf{13}, 16317.

\bibitem{56} K. Maki, \textit{Physical Review}, 1966, \textbf{148}, 362.

\bibitem{57} R. Chen, L. Li, L. Zeng, K. Li, P. Yu, K. Wang, Z. Xiang, S. Wang, J. Qin and W. Zhang \textit{et al.}, \textit{Acta Materialia}, 2025, \textbf{298}, 121401.
\bibitem{58} E. F. Talantsev, \textit{Review of Scientific Instruments}, 2022, \textbf{93}, 053912.

\bibitem{59} C. Kittel and P. McEuen, \textit{Introduction to Solid State Physics}, John Wiley \& Sons, 2018.

\bibitem{60} J. Carbotte, \textit{Reviews of Modern Physics}, 1990, \textbf{62}, 1027.

\bibitem{61} W. McMillan, \textit{Physical Review}, 1968, \textbf{167}, 331.

\bibitem{62} C. P. Bean, \textit{Reviews of Modern Physics}, 1964, \textbf{36}, 31.

\bibitem{63} J. H. Kim, R. Hidayati, S.-G. Jung, Y. A. Salawu, H. J. Kim, J. H. Yun and J.-S. Rhyee, \textit{Acta Materialia}, 2022, \textbf{232}, 117971.

\bibitem{64} S. J. Singh, M. Bristow, W. R. Meier, P. Taylor, S. J. Blundell, P. C. Canfield and A. I. Coldea, \textit{Physical Review Materials}, 2018, \textbf{2}, 074802.

\bibitem{65} M. J. Rice, \textit{Physical Review Letters}, 1968, \textbf{20}, 1439--1441.

\bibitem{66} K. Kadowaki and S. B. Woods, \textit{Solid State Communications}, 1986, \textbf{58}, 507--509.

\bibitem{67} B. T. Matthias, \textit{Physical Review}, 1955, \textbf{97}, 74.

\bibitem{68} M. Collver and R. Hammond, \textit{Physical Review Letters}, 1973, \textbf{30}, 92.

\bibitem{69} M. Capone, M. Fabrizio, C. Castellani and E. Tosatti, \textit{Science}, 2002, \textbf{296}, 2364--2366.

\bibitem{70} O. N. Senkov and D. B. Miracle, \textit{Materials Research Bulletin}, 2001, \textbf{36}, 2183--2198.

\bibitem{71} H. Song, F. Tian, Q.-M. Hu, L. Vitos, Y. Wang, J. Shen and N. Chen, \textit{Physical Review Materials}, 2017, \textbf{1}, 023404.

\bibitem{72} H. Wang, P. Yang, W. Zhao, S. Ma, J. Hou, Q. He, C. Wu, H. Chen, Q. Wang and Q. Cheng \textit{et al.}, \textit{Nature Communications}, 2024, \textbf{15}, 6782.

\bibitem{73} C. Lee, G. Song, M. C. Gao, R. Feng, P. Chen, J. Brechtl, Y. Chen, K. An, W. Guo and J. D. Poplawsky \textit{et al.}, \textit{Acta Materialia}, 2018, \textbf{160}, 158--172.

\bibitem{74} K. Jasiewicz, J. Tobola and B. Wiendlocha, \textit{Physical Review B}, 2023, \textbf{108}, 224505.
\bibitem{75} D. H. Cook, X. Chen, M. O’Leary, S. Rao, W. Wang, C. H. Belcher, M. Payne, P. Borges, T. L. Lee, A. Devaraj, et al., \textit{Nature}, (2026), \textbf{1--7}.
\end{thebibliography}
\end{document}